

\magnification=\magstep1

\parskip=3pt plus1pt minus.5pt

\def\half{{\textstyle {1 \over 2}}}

\def\quarter{{\textstyle {1 \over 4}}}

\def\frac#1#2{{{#1}\over{#2}}}

\def\edth{\hbox{$\partial$\kern-0.4em\raise0.03ex\hbox{\'{}}
\kern-0.3em\hbox{}}}


\def\sqr#1#2{{\vcenter{\vbox{\hrule height.#2pt
      \hbox{\vrule width.#2pt height#1pt \kern#1pt
         \vrule width.#2pt}
      \hrule height.#2pt}}}}

\outer\def\exclaim #1. #2\par{\medbreak
  \noindent{\sl#1.\enspace}{\rm#2}\par
  \ifdim\lastskip<\medskipamount \removelastskip\penalty55\medskip\fi}

\outer\def\proof #1\par{\medbreak
  \noindent{\sl Proof.\enspace}{\rm#1}\par
  \ifdim\lastskip<\medskipamount \removelastskip\penalty55\medskip\fi}


\def\({\left(}
\def\){\right)}
\def\<{\left\langle}
\def\>{\right\rangle}

\def\[{\left[}
\def\]{\right]}

\let\text=\hbox

\font\titlefont=cmcsc10 scaled \magstep2

\centerline {\titlefont Naked And Thunderbolt Singularities}
\centerline{ \titlefont In Black Hole Evaporation}
\vskip 0.6truein
\centerline {S. W. Hawking}
\centerline{ \&}
\centerline{ J. M. Stewart}

\vskip .5truein
\centerline {Department of Applied Mathematics and Theoretical
Physics}
\centerline {University of Cambridge}
\centerline {Silver Street}
\centerline {Cambridge CB3 9EW}
\centerline {UK}
\vskip .5truein
\centerline {\it July 1992 }
\vskip 0.7truein
\centerline {\bf Abstract}
 \bigskip
If an evaporating black hole does not settle down to a non radiating
remnant, a description by a semi classical Lorentz metric must contain
either a naked singularity or what we call a thunderbolt, a
singularity that spreads out to infinity on a spacelike or null path.
We investigate this question in the context of various two dimensional
models that have been proposed. We find that if the semi classical
equations have an extra symmetry that make them solvable in closed
form, they seem to predict naked singularities but numerical
calculations indicate that more general semi classical equations, such
as the original CGHS ones give rise  to thunderbolts. We therefore
expect that the semi classical approximation in four dimensions will
lead to thunderbolts. We interpret the prediction of thunderbolts as
indicating that the semi classical approximation breaks down at the
end point of black hole evaporation, and we would expect that a full
quantum treatment would replace the thunderbolt with a burst of high
energy particles. The energy in such a burst would be too small to
account for the observed gamma ray bursts.

 \vfill \eject

\beginsection 1 Introduction

It has been known for some time that classical general relativity
predicts singularities in gravitational collapse. At the
singularities, the Einstein equations will not be defined. Thus there
will be a limit as to how far into the future one can predict
spacetime. However, it seems that singularities formed in
gravitational collapse always occur in regions that are hidden from
infinity by an event
horizon, so the breakdown of the Einstein equations at the singularity
does not affect our ability to predict the future in the asymptotic
region of space. This assumption that the singularities are hidden is
known as the Cosmic Censorship Hypothesis and is fundamental to all
the work that has been done on black holes. It remains unproven  but
it is almost certainly true for classical general relativity
with a suitable definition of a singularity that is so bad it can't be
smoothed out or continued through.

On the other hand, in the semi classical approximation to quantum
gravity a black hole formed in a  gravitational collapse will  emit
thermal radiation and evaporate slowly. If the black hole has a charge
that is coupled to a long range field and which can't be radiated,
such as a magnetic charge, it may be able to settle down to a non
radiating state such as the extreme Reissner-Nordstr\o m solution.
But for black holes without such a charge, there are no zero
temperature classical solutions they can settle down to. One might
suppose they settled down to some stable or semi stable remnant that
was not a classical solution but was maintained by quantum effects.
However, quite apart from the fact that there is nothing very obvious
to stabilize such remnants, their existence would create severe
problems.
If they had a mass of the order of the Planck mass, one might have
expected that there would be more than the cosmological critical
density of the remains of black holes formed in the very early
universe. While if they had zero mass, they would lead to infinite
degeneracy of the vacuum state.

The most natural assumption would seem to be that black holes without
a conserved charge disappear completely. To suppose that black holes
could be formed but never disappear would violate CPT unless there
were also a separate species of white holes which would  have existed
from the beginning of the universe. On the other hand, if black holes
disappear completely, black and white holes can be different
aspects of the same objects, which would be an aesthetically
satisfying solution to the CPT problem. Holes would be called black
when they were large and classical, and not radiating much, but they
would be called white when the quantum emission was the dominant
process.

If black holes disappear completely, this can not be described by a
Lorentzian metric without some sort of naked singularity, or what
would be even worse, a region of closed time like curves. Spreading
out from the naked singularity or region of chronology violation would
be a Cauchy horizon. Beyond this horizon the semi classical equations
would not uniquely specify  the solution, but one would hope
that it would determined by a full quantum treatment, though maybe
with loss of quantum coherence. Otherwise, we could be in for a
surprise every time a black hole on our past light cone evaporates.

Within the context of the semi classical approximation there is
however an alternative to a naked singularity that has not received
much attention. We shall call it a thunderbolt. It is a singularity
that spreads out to infinity on a space like or null path.  It is not
a naked singularity because you don't see it coming until it hits you
and wipes you out. It would mean that the semi classical equations
could not only not be evolved uniquely (as with a naked singularity),
but they could not be evolved at all more than a finite distance into
the future. If the thunderbolt was null, one could regard it as the
singular Cauchy horizon produced by  some would-be naked singularity.
This would be like what is believed to happen to the inner Cauchy
horizons of classical black holes under generic perturbations.
One might therefore expect that although the semi classical equations
could lead to naked singularities in special situations, one would get
a thunderbolt if one perturbed the equations or the initial data
slightly.

If the semi classical equations were to predict a thunderbolt
singularity as the end point of black hole evaporation, one would have
to conclude that the singularity would be softened and smeared out by
quantum effects because surely many black holes must have evaporated
in the past, and yet we survived. Nevertheless, if the semi classical
equations predict thunderbolts, this might indicate that
something fairly dramatic happens in the full quantum theory.

In four dimensions, the one loop corrections are quadratic in the
curvature. This means that the semi classical equations including one
loop back reaction are fourth order and have unphysical runaway
solutions. It is therefore hard to use them to decide whether the
evaporation of black holes leads to naked singularities or
thunderbolts. On the other hand, the the one loop corrections in two
dimensions
are proportional to the curvature scalar. This means that the semi
classical equations are second order even when the back reaction is
taken into account. It should therefore be possible to decide what
they predict as the outcome of black hole evaporation. Hopefully, this
will give an indication of what might happen in four dimensions.

In two dimensions the Einstein Hilbert Lagrangian $ R $ is a
divergence. This means that to get a non trivial interaction with the
metric, one has to multiply the Einstein Hilbert term by a function of
a dilaton field $\phi $. An interesting model in which the metric is
coupled to a dilaton field and $N$ minimal scalars has been proposed
by Callan, Giddings, Harvey and Strominger [1], (henceforth referred
to as CGHS). In the classical
version of this theory one can form a black hole by sending in a wave
of one of the scalar fields from the asymptotic region. Quantum field
theory on this classical black hole background then shows that the
black hole will radiate thermally in each of the fields. Presumably
this means that the black holes will evaporate but a full quantum
treatment of the problem seems too difficult even in this simple
theory. However, Callan et al suggested that in the large $N$ limit,
one could neglect ghosts and quantum fluctuations of the metric and
dilaton in comparison with those of the scalar fields. The effective
action arising from the scalar quantum loops would be completely
determined by the trace anomaly and the conservation equations
together with boundary conditions. One could therefore add it to the
classical action for the metric and dilaton fields and obtain a set of
semi classical hyperbolic differential equations for the metric and
dilaton.

Even these relatively simple equations have not been solved in closed
form. Callan et al hoped that the result of including the action of
the scalar loops would be to cause a black hole to evaporate
completely without any singularity and tend at late times to the
linear dilaton solution, which is the analogue of Minkowski space, and
which is the natural candidate for a ground state. However, later
work showed that there was necessarily a singularity, and that the
solution could not settle down to a static state in which the
singularity remained hidden behind an event horizon.

These results presumably indicate that the semi classical equations
lead either to a naked singularity or a thunderbolt. But which? The
original semi classical equations proposed by CGHS do not seem to
admit closed form solutions. Various authors have suggests
modifications to the semi classical equations that introduce an extra
symmetry and make the equations solvable in closed form. We shall show
the exact solutions have naked singularities. However they also
continue to emit radiation at a finite rate and the mass becomes
arbitrarily negative. Such behaviour is presumably unphysical, or at
least one hopes so. The conservation of energy would lose its
practical significance if one could have negative mass naked
singularities. In one case at least, one could use the non uniqueness
of the solution
after the naked singularity has appeared to cut off the analytically
continued exact solution at the Cauchy horizon produced by the naked
singularity and glue on a non radiating solution. This procedure
however transforms the Cauchy horizon into a thunderbolt singularity,
although a fairly mild one.

In the four dimensional case, the equations don't have symmetries that
allow one to solve them in closed form. There is thus no reason to
expect special properties like conformal symmetry in two dimensional
models of black holes. We shall therefore investigate the behaviour of
solutions of the original semi classical equations proposed by CGHS
which we expect to be more typical of the general case.
Since these equations do not admit solutions in closed form, there
seems no alternative but to integrate the equations numerically.
Fortunately hyperbolic equations in 1+1 dimensions are relatively easy
and there are reliable and numerically stable routines available. To
test their accuracy, we first applied them to the equations without
back reaction. We obtained excellent agreement with the known
solution, the Witten two dimensional black hole. Encouraged by this,
we included the back reaction terms and obtained results that strongly
indicate a thunderbolt. This supports our view that while naked
singularities may occur for certain sets of semi classical equations
with special symmetries, more general two dimensional models of black
hole evaporation will exhibit thunderbolts.

In section 2 the model and the various  sets of semi classical
equations are described. Those with special symmetries that allow
exact solutions are shown to lead to naked singularities in section 3
while in section 4  the numerical results of integrating more general
equations are presented. A test is given to distinguish a thunderbolt
from an eternal black hole. The implications for black holes
in four dimensions are discussed in section 5. The numerical algorithm
used is described in an appendix.

\beginsection 2.  The semi classical model

CGHS assume the spacetime contains a dilaton field $\phi $ and $N$
minimally coupled scalar fields $f_i$, described by a classical
Lagrangian
$$
L = \frac{1}{2\pi}\sqrt{-g}\[
e^{-2\phi }\(R + 4(\nabla \phi )^2 + 4\lambda ^2\) -
\frac{1}{2} \sum_{i=1}^N (\nabla f_i)^2\]\, \eqno(2.1)
$$
where $R$ is the Ricci scalar and $\lambda $ is a coupling constant.

Any two dimensional spacetime is of course conformally flat, so one
can introduce null coordinates $x^\pm$ and write the line element as
$$
ds^2 = -e^{2\rho}\,dx^+\,dx^-\;. \eqno(2.2)
$$
CGHS suggested that in the limit of a large number $N$ of scalar
fields $f_i$ one could neglect the quantum fluctuations of the dilaton
and the metric, and treat the back reaction in the scalar fields
semi classically by adding to the action a trace anomaly term
$$
- \kappa \partial_+\rho \,\partial_-\rho. \eqno(2.3)
$$
CGHS took $\kappa = N/12$. However taking ghosts into account
leads to
$$
\kappa ={ N -24\over 12}, \eqno(2.4)
$$
in that theory.
For consistency with refs [3--5] we henceforth define $\kappa $ by
(2.4).
Occasionally we shall use the earlier value in the form $\tilde\kappa
=N/12$; obviously $\tilde\kappa =\kappa +2$.
We shall call the
theory defined by equations (2.1), (2.3) and (2.4) the original
theory.

Strominger [2] has suggested that the ghosts should be coupled
to
a different metric. This leads to the action of the original theory
(with $\kappa $ replaced by $\tilde\kappa $), plus an additional term
$$ 2\( \partial_+ \phi \,\partial_-\phi -
\partial_+\phi \,\partial_-\rho -
\partial _+ \rho \,\partial _- \phi+
\partial_+\rho\,\partial_-\rho\). \eqno( 2.5)
$$
We
shall call this the decoupled ghost theory, though in fact the ghosts
are still coupled to the geometry, only differently.

de Alwis [3] and Bilal and Callan [4]  have suggested that the
cosmological
constant $\lambda \sp 2$ term be multiplied by a function $D(\phi)$ to
make the theory conformally invariant where
$$
D(\phi)= {1\over 4}(1+y)^2\exp \left[ {1-y\over 1+y}\right]\eqno(2.6)
$$
and $y=\sqrt{1- \kappa e^{2 \phi}} $. We shall call this the conformal
theory. It can be solved in closed form.

Another Lagrangian with a special symmetry that has a conserved
current $ j \sp \mu = \partial _{\mu}( \phi - \rho)$ has been proposed
by Russo, Susskind and Thorlacius [5]. It is the Lagrangian of the
original theory plus the additional term
$$ - \kappa \phi \,\partial _+ \partial _- \rho \eqno( 2.7)
$$
We shall call this the conserved current theory.

The general solution of the conformal and conserved current theories
with an asymptotically flat weak coupling region will be given in
section 3. It will be shown they have naked singularities for positive
$\kappa $. Here we give the field equations for the two Lagrangians
without special symmetries, the original and decoupled ghost theories.
The evolution equations can be written in the form
$$
\eqalignno {
\partial_+ \partial_- f_i = &0, &(2.8a)\cr
\partial_+ \partial_- \rho = &P^{-1} (2\partial_+\phi\,\partial_-\phi
+
Y), &(2.8b) \cr
\partial_+\partial_-\phi = & Q \partial_+\partial_-\rho, &(2.8c)\cr
}
$$
where we have introduced the quantities
$$
\eqalign {
P= & 1- \kappa e^{2\phi} \cr
Q= &1-\half \kappa e^{2\phi}, \cr
}\eqno(2.9a)
$$
in the original theory, and
$$
\eqalign {
P=&1-\tilde\kappa e^{2\phi}+ \half \tilde\kappa e^{4\phi} \cr
Q=&1- \half \tilde\kappa e^{2\phi}, \cr
} \eqno(2.9b)
$$
in the decoupled ghost theory.
Here
$$
Y = \frac{1}{2}\lambda ^2 e^{2\rho}. \eqno(2.10)
$$
In addition  there are two constraint equations.
In the original theory they are
$$
\eqalignno {
e^{-2\phi}\(2\partial_+^2\phi - 4\partial_+\phi\,\partial_+\rho\) -
\kappa \[\partial_+^2\rho - (\partial_+\rho)^2 - t_+(x^+)\] =
& \half \sum _i(\partial_+ f_i)^2, & (2.11a)\cr
e^{-2\phi}\(2\partial_-^2\phi - 4\partial_-\phi\,\partial_-\rho\) -
\kappa \[\partial_-^2\rho - (\partial_-\rho)^2 - t_-(x^-)\] =
& \half \sum _i(\partial_- f_i)^2, & (2.11b)\cr
}
$$
where $t_\pm$ are arbitrary functions.
They are constraints in the following sense.
(2.11a,b) need be imposed only on surfaces $x^- = const.$ and
$x^+ = const.$ respectively.
They hold then throughout the spacetime as a consequence of the
evolution equations.
The constraints for the decoupled ghost theory involve replacing
$\kappa $ by $\tilde\kappa $ as well as adding some extra
terms which vanish when $\phi=\rho$.
Since we only impose the constraints on the initial surfaces where we
may also set $\phi=\rho$ (see later), we do not need to write down
explicitly the constraints for this theory.
One may easily recover the classical equations, i.e., without the
trace anomaly term, by setting $\kappa =0$ in the equations of the
original theory.

We consider first solutions of the classical equations.
Equations (2.8b,c) have the solution
$$
e^{-2\phi } = e^{-2\rho} = \frac{M}{\lambda } - \lambda ^2 x^+ x^-,
\eqno(2.12)
$$
where $M$ is a constant, and arbitrary additive constants to $x^\pm$
have been ignored.
If $M=0$ we obtain the so-called linear dilaton, but if $M\ne 0$
the solution represents a black hole of mass $M$, with horizons given
by $x^+ x^- = 0$ and a singularity when $x^+ x^- = M/\lambda ^3$.

Consider next the situation where a linear dilaton occurs for $x^+ <
x^+_o$.
At $x^+_o$ a matter wave described by $f = f(x^+)$, which is a
solution
of (2.8a) propagates in the $x^-$-direction.
If $f(x^+)$ has compact support, then once the wave has passed, the
spacetime will once again be described by (2.12), but now we must
expect $M\ne0$.
For simplicity we consider an impulsive wave described by
$$
\half \sum _i(\partial_+ f_i)^2 = a\delta (x^+ - x^+_o), \eqno(2.13)
$$
where $a$ is a constant.

We now apply the constraint (2.11a) on an initial surface
$x^- = x^-_o$.
On such a surface the value of $\rho$ is arbitrary; changes correspond
to a rescaling of coordinates.
We may therefore choose $\rho=\phi $ on this surface.
Then (2.11a), (2.13) imply a jump increase in $\partial _+(e^{-2\phi
})$ at
$x^+ = x^+_o$, i.e.,
$$
e^{-2\phi } = e^{-2\rho} =
ax^+_o - \lambda ^2\(x^-_o + \frac{a}{\lambda ^2}\)x^+ \eqno(2.14)
$$
for $x^+ \ge x^+_o$ on $x^- = x^-_o$.
Comparing this data with (2.12) we see that we have a black hole
solution
$$
e^{-2\phi } = e^{-2\rho} = ax^+_o - \lambda ^2
\(x^- + \frac{a}{\lambda ^2} \) x^+, \eqno(2.15)
$$
for $x+ > x^+_o$.
An alternative, more long-winded approach (in this case) is to solve
(2.8b,c) as a characteristic initial value problem.
If $\rho$ and $\phi $ are specified for $x^+ \ge x^+_o$ on $x^- =
x^-_o$, and for $x^- \ge x^-_o$ on $x^+ = x^+_o$, then the solution
is determined locally and uniquely for $x^\pm \ge x^{\pm o}$.
One piece of the data is given by (2.14).
The other follows from continuity at $x^+ = x^+_o$, viz.
$$
e^{-2\phi } = e^{-2\rho} = -\lambda ^2 x^+_ox^-, \eqno(2.16)
$$
for $x^- \ge x^{_-o}$ on $x^+ = x^+_o$.

We turn now to the semi classical analogue.
For $x^+ < x^+_o$ the classical solution is
$$
e^{-2\phi } = e^{-2\rho} = -\lambda ^2 x^+x^-. \eqno(2.17)
$$
This is also a solution of the semi classical equations.
(Both sides of (2.8b) vanish for arbitrary $\kappa $.)
After the shock we have to apply the constraint (2.11a) for
$x^+ > x^+_o$ on $x^- = x^-_o$.
We still have the coordinate freedom to choose $\rho = \phi $ on this
surface. We wish to study the situation in which there is no incoming
energy momentum apart from the matter wave. This corresponds to
choosing $t_+(x^+)$ so that the factor multiplying $\kappa $ in
(2.11a) vanishes on $ x^-=x^-_o$. Similarly, we want no energy
momentum coming from the linear dilaton region. This corresponds to
choosing
$ t_-(x^-)$ so that the term multiplying $\kappa $ in (2.11b) is zero.
With this choice equations
(2.14) and (2.16) form characteristic initial data for the
semi classical evolution equations (2.8b,c).
However we lack an exact solution to these equations in the absence of
some special symmetry. We shall therefore resort to numerical
integration in section 4.

\beginsection 3.  The conformal and conserved current theories

In the conformal theory of de Alwis and Bilal \& Callan the
Lagrangian may be written as
$$
\eqalign {
L = &\frac{1}{\pi}\bigg[ e^{-2\phi }\(2\partial_+\rho\,\partial_-\phi
+
2\partial_-\rho\,\partial_+\phi -4\partial_+\phi \,\partial_-\phi \) +
\cr
& \qquad \frac{1}{2}\sum_{i=1}^N \partial_+ f_i \,\partial_- f_i -
\kappa \partial_+\rho\,\partial_-\rho +
\lambda ^2e^{2\rho-2\phi }D(\phi ) \bigg],\cr}
\eqno(3.1)
$$
where $D(\phi )$ was given earlier by equation (2.6).
Bilal and Callan suggested a sequence of changes of dependent
variables
$$
\omega = \frac{1}{\sqrt{|\kappa|} e^\phi }, \qquad
\chi = \half(\rho + \epsilon \omega ^2), \eqno(3.2a)
$$
where $\epsilon = \kappa/|\kappa |$, followed by
$$
\Omega = \half\epsilon \omega \sqrt{\omega ^2 - \epsilon } -
\half \log(\omega +\sqrt{\omega ^2 - \epsilon }). \eqno(3.2b)
$$
These produce a free field Lagrangian
$$
L = \frac{1}{\pi}\bigg[4\kappa \partial_+\Omega \,\partial_-\Omega -
4\kappa \partial_+\chi\,\partial_-\chi +
\frac{1}{2}\sum_{i=1}^N \partial_+ f_i \,\partial_- f_i +
\lambda ^2 e^{2\rho-2\phi }D(\phi )\bigg],\eqno(3.3)
$$
and constraints
$$
2\kappa {\partial_\pm}^2\chi  +
4\kappa \partial_\pm\Omega \,\partial_\pm\Omega -
4\kappa\partial_\pm\chi \,\partial_\pm\chi +
\frac{1}{2}\sum_{i=1}^N \partial_\pm f_i \,\partial_\pm f_i +
t_\pm(\sigma^\pm) = 0. \eqno(3.4)
$$
Note that Bilal and Callan used rescaled asymptotically Minkowskian
coordinates
$$
\sigma^+ = \log(x^+), \qquad \sigma^- = -\log(-x^-),
$$
where $x^\pm$ are the coordinates used in section 2.
\medskip
The equations of motion simplify after a further change of dependent
variables $\Psi _\pm = \chi \pm \Omega $  to
$$
\partial_+\partial_-\Psi _- = 0, \qquad
\partial_+\partial_-\Psi _+ = -\frac{\lambda ^2}{4e}e^{4\Psi _-}.
\eqno(3.5)
$$
The first equation is the standard wave equation in characteristic
coordinates, and the second is similar but with a known source term.
Bilal and Callan wrote the solution in the form
$$
\eqalign {
2\Psi _- = & \alpha (\sigma^+) + \beta (\sigma^-) + K ,\cr
2\Psi _+ = & 2\gamma (\sigma^+) + 2\delta (\sigma^-) -
\alpha (\sigma^+) - \beta (\sigma^-) + K -
\frac{2}{\kappa }\int ^{\sigma^+} e^{2\alpha (s)}\,ds\;
\int^{\sigma^-} e^{2\beta (t)}\,dt,\cr
} \eqno(3.6)
$$
where
$$
K = \frac{1}{2} + \log\(\frac{2}{\lambda \sqrt{|\kappa |}}\),
$$
and $\alpha $, $\beta $, $\gamma $ and $\delta $ are arbitrary
functions of one variable, to be determined from the initial data
and constraints.
Bilal and Callan chose to take $t_\pm(\sigma^\pm)=0$ in (3.4), which
can be rewritten as
$$
\eqalign{
(\partial_+\gamma)^2 - (\partial_+(\gamma-\alpha))^2
-{\partial_+}^2\gamma
= & \frac{1}{2\kappa}\sum_{i=1}^N(\partial_+ f_i)^2,\cr
(\partial_-\delta)^2 - (\partial_-(\delta-\beta))^2 -{\partial_-
}^2\delta
= & \frac{1}{2\kappa}\sum_{i=1}^N(\partial_- f_i)^2.\cr
}\eqno (3.7)
$$
\medskip
The linear dilaton is not a solution of this theory.
Consider however  static solutions, i.e., depending on
$\sigma = \half(\sigma^+ - \sigma^-)$ only, which are asymptotic to
the linear dilaton as $\sigma\rightarrow\infty$.
Bilal and Callan obtained the general solution in the form
$$
\eqalign{
\alpha(\sigma^+) = \half\sigma^+, \qquad
\gamma(\sigma^+) = & \quarter\sigma^+ + \half T +
\half\log\frac{|\kappa|}{4e},\cr
\beta(\sigma^-) = -\half\sigma^-, \qquad
\delta(\sigma^-) = & -\quarter\sigma^- + \half T,\cr
} \eqno (3.8)
$$
where $T$ is a constant that behaves like the mass.
\medskip
Bilal and Callan modelled the shockwave problem by requiring the
solution (3.8) to hold for $\sigma^+ < 0$ and setting
$\half\sum(\partial_-f_i)^2=0$,
$\half\sum(\partial_+f_i)^2 = a\delta(\sigma^+)$ in (3.7).
The solution is
$$
\eqalign{
\alpha(\sigma^+) = \half\sigma^+, \qquad
\gamma(\sigma^+) = & \quarter\sigma^+ + \half T -
\frac{a}{\kappa}(e^{\sigma^+}-1)\theta(\sigma^+) +
\half\log\frac{|\kappa|}{4e},\cr
\beta(\sigma^-) = -\half\sigma^-, \qquad
\delta(\sigma^-) = & -\quarter\sigma^- + \half T,\cr
} \eqno (3.8)
$$
and (3.6) now implies that for $\sigma^+ > 0$
$$
\eqalign{
2\Omega(\phi) =& \frac{1}{\kappa}e^{\sigma^+-\sigma^-}-
\frac{a}{\kappa}(e^{\sigma^+}-1) - \frac{1}{4}(\sigma^+ - \sigma^-) +
T + \frac{1}{2}\log\frac{|\kappa|}{4e},\cr
\rho + \log\lambda + \frac{1}{\kappa} e^{-2\phi}= &
\frac{1}{\kappa}e^{\sigma^+-\sigma^-}-
\frac{a}{\kappa}(e^{\sigma^+}-1) - \frac{1}{4}(\sigma^+ - \sigma^-) +
T.\cr
}\eqno(3.9)
$$
\medskip
The variables $\chi$ and $\Omega$ are regular functions of position.
However there may be a singularity where the curvature scalar
$R=8e^{-2\rho}\partial_+\partial_-\rho$ diverges.
In order to locate this recall from (3.2) that $\rho = 2\chi -
\epsilon\omega^2$ and $\Omega = \Omega(\omega)$.
Then
$$
\partial_+\partial_-\rho = 2\partial_+\partial_-\chi -
2\frac{\epsilon\omega}{\Omega'} \partial_+\partial_-\Omega -
\frac{2\epsilon}{{\Omega'}^2}\(1-\omega\frac{\Omega''}{\Omega'}\)
\partial_+\Omega\,\partial_-\Omega.
$$
Thus the singularity occurs when $\Omega' = 0$.
However from (3.2b) we see that this occurs when $\omega^2 = \epsilon$
or $\Omega = 0$ and only for $\kappa > 0$.
The apparent horizon is located where $\partial_+\phi = 0$ or
equivalently
where $\partial_+\Omega = 0$.

We now demonstrate that in this theory the singularity is eventually
naked, i.e., the apparent horizon moves to the future of the
singularity.
Let the singularity be located at $\sigma^- =\sigma_s^-(\sigma^+)$,
and the apparent horizon at $\sigma^- = \sigma_h^-(\sigma^+)$.
If $T$ is positive, the singularity will start off inside the apparent
horizon. However for large $\sigma^+$ (3.9) implies
$$
\eqalign{
\sigma^-_s \sim & -\log(a+\quarter\kappa\sigma^+e^{-\sigma^+}),\cr
\sigma^-_h \sim & -\log(a+\quarter\kappa e^{-\sigma^+}).\cr
} \eqno (3.10)
$$
In this limit $\sigma^-_s$ approaches $\sigma^-_h$ from below, i.e.,
the apparent horizon and the singularity are ultimately tangent with
the singularity to the past of the horizon. So the apparent horizon
and singularity must meet and cross at some point $(\sigma
^+_n,~\sigma
^+_n)$. See figure 1. At this point the singularity will become
timelike and naked. The line $\sigma ^-= \sigma ^-_n, ~\sigma ^+>
\sigma ^+_n$ will become a Cauchy horizon. Although the exact solution
continues smoothly beyond the Cauchy horizon, it is unphysical
because it has a steady outflow of radiation and an effective mass
parameter that becomes arbitrarily negative. It should be noted that
the argument does not depend on the impulsive
nature of the shock, and can be generalized easily to
arbitrary but finite infalls of matter.
\bigskip
The analysis of the conserved current theory of Russo, Susskind and
Thorlacius is very similar.
They use the coordinates $x^\pm$ of section 2 and auxiliary variables
$$
\Omega = \frac{\sqrt\kappa}{2}\phi +
\frac{e^{-2\phi}}{\sqrt\kappa}, \qquad
\chi = \sqrt{\kappa}(\rho - \phi) + \Omega .\eqno(3.11)
$$
The Lagrangian is
$$
S = \frac{1}{\pi}
\[\partial_+\Omega\,\partial_-\Omega - \partial_+\chi\,\partial_-\chi
+
\lambda^2e^{2(\chi-\Omega)/\sqrt{\kappa}} -
\frac{1}{2}\sum_{i=1}^N\partial_+ f_i\,\partial_- f_i\], \eqno (3.12)
$$
with constraints
$$
\sqrt{\kappa}\partial_\pm^2\chi - \partial_\pm\chi\,\partial_\pm\chi +
\partial_\pm\Omega\,\partial_\pm\Omega +
\frac{1}{2}\sum_{i=1}^N\partial_\pm f_i\,\partial_\pm f_i -
\kappa t_\pm(x^\pm) = 0. \eqno (3.13)
$$
\medskip
The asymptotically flat static geometries with $\phi=\rho$ are given
by
$$
\Omega=\chi=-\,\frac{\lambda^2 x^+ x^-}{\sqrt{\kappa}} +
P\sqrt{\kappa}\log(-\lambda^2x^+x^-) +
\frac{M}{\lambda\sqrt{\kappa}}, \eqno (3.14)
$$
where $P$ and $M$ are constants.
Setting $P=-\quarter$ and $M=0$ gives the linear dilaton vacuum.
Russo et al  constructed a solution which is the dilaton
for $x^+ < x^+_o$ and corresponds to infalling matter
for $x^+ > x^+_o$ with
$$
\frac{1}{2}\sum_{i=1}^N(\partial_+ f_i)^2 =
 a \delta(x^+ - x^+_o),
$$
viz.
$$
\Omega=\chi=-\,\frac{\lambda^2 x^+x^-}{\sqrt \kappa} -
\frac{\sqrt\kappa}{4}\log(-\lambda^2 x^+x^-) -
\frac{a }{\sqrt{\kappa}}(x^+-x^+_o)\theta(x^+ - x^+_o).
\eqno(3.15)
$$
The singularity is again given (for $\kappa > 0$) by $\Omega'=0$ which
implies $\Omega = \quarter\sqrt{\kappa}(1 - \log\quarter\kappa)$.
Again the singularity starts off inside the apparent horizon. However
for large $x^+$, the singularity will be at
$$
x^-_s \sim -\,\frac{a }{\lambda^2 } -
\frac{\kappa}{4\lambda^2}\frac{\log x^+}{x^+},
$$
while the apparent horizon is located at
$$
x^-_h = -\,\frac{a}{\lambda^2} -
\frac{\kappa}{4\lambda^2}\frac{1}{x^+}.
$$
Again the singularity and apparent horizon must meet and cross at a
point $(x^+_n, ~x^-_n)$ where the singularity will become timelike and
naked. In this case one can cut off the solution  (3.15) on the Cauchy
horizon $x^-=x^-_n$ and join on to the linear dilaton solution. This
makes the Cauchy horizon a mild thunderbolt singularity.

\beginsection 4. Numerical results for theories without special
symmetries

In both the classical and semi classical problems a numerical
treatment is not entirely straightforward, for singularities are
present in each.
Fortunately we know the analytic solution for the former.
We have therefore developed a numerical algorithm which handles the
classical problem in a satisfactory manner, and have then applied it
to the semi classical case.

The parameter $\lambda $ may be scaled away, and so we have chosen
$\lambda =1$.
The values $x^\pm_o$ defining the initial data surfaces are
arbitrary, and we have chosen $x^\pm_o = \pm 1$, except where stated
otherwise.
{}From (2.15) we see that the black hole singularity occurs on the
hyperbola
$$
x^- = \frac{a}{\lambda ^2}\(\frac{x^+_o}{x^+}-1\), \eqno(4.1)
$$
and the relevant apparent horizon is at
$$
x^- = -\frac{a}{\lambda ^2}. \eqno(4.2)
$$
Thus by choosing $a\in (0,1)$ we may ensure that the singularity
remains in the domain of dependence of our data.
The graphs are all drawn for $a=0.9$, thought to be a generic case.

At the singularity the Ricci curvature scalar
$R = 8e^{-2\rho}\partial_+\partial_-\rho$ becomes singular.
Figure~2 shows $\arctan R$ as a function of $x^\pm$ for the classical
case.
The solution is not defined to the future of the singularity (top
right of the surface) but for convenience in drawing the surface $R$
has been assigned a token value of $\infty$ in this region.
Also shown is the apparent horizon $x^- = -a$.
For large $x^+$ the singularity approaches the apparent horizon as
predicted by (4.1,2).

The same computational algorithm was adopted for the semi classical
equations.
{}From (2.8b) $R$ may be expected to become singular once $\phi $ has
increased to a value $\phi _c$ at which $P=0$.
Such a $\phi_c$ exists for the original theory if $\kappa > 0$ and for
the decoupled ghost theory if $\tilde\kappa > 2$.
We have also considered $ \tilde\kappa <2$ that is $N<24$ for the
decoupled ghost theory.
For each fixed $x^+$ the programme integrated the equations in the
direction of increasing $x^-$.
The singularity was deemed to occur at the point where $\phi =\phi _c$
and the apparent horizon when $\partial_+\phi $ changed sign.
For $\kappa >0$, the results do not seem to depend sensitively on the
exact value so $\kappa =0.5$ or $0.8$ was used for the graphs
presented here.
The behaviour shown in Figure~3 for the original theory
looked at first sight broadly similar to the black hole case,
although the singularity is a little steeper.
However the two solutions are radically different. To show this we
need a test that will distinguish a black hole singularity that
remains at a fixed position from a thunderbolt that spreads out to
infinity.

Consider the outgoing null geodesic $x^-=x^-_o$, with tangent vector
$T^+ = dx^+/dt$ where $t$ is an affine parameter.
For each fixed value of $x^+$ we consider the ingoing null geodesic
$x^+=const.$ with tangent vector $T^-=dx^-/ds$ and affine parameter
$s$ normalized by say $g(T^+, T^-) = \half$ at $(x^+, x^-_o)$.
The affine parameter distance to the horizon will be denoted $s(x^+)$.
For a black hole we would expect that for large $x^+$, $s(x^+)$ will
be asymptotically a linear function of $t(x^+)$, c.f.,
Schwarzschild.
Indeed it is an elementary exercise to carry out the calculations
analytically for the black hole in the classical shockwave problem,
finding
$$
\eqalign {
t(x^+) = &-(a+x^-_o)^{-1}\log\(ax^+_o-(a+x^-_o)x^+\),\cr
s(x^+) = &(-ax^+_o/x^+ + a + x^-_o)\[\log(ax^+_o) +
(a+x^+_o)t(x^+)\].\cr
} \eqno(4.3)
$$
As $x^+ \rightarrow \infty $, $t\rightarrow \infty $ and
$s(x^+) \sim (a+x^-_o)^2 t $, giving linear behaviour.

We next developed a numerical algorithm to explore the behaviour of
$s(t)$ for large $t$.
Numerically this is not entirely straightforward.
Firstly one wants to explore very large values of $x^+$, i.e., to
integrate over an enormous number of grid points.
Secondly the singularity is approaching the horizon asymptotically,
c.f., (4.1,2)!
Figure~4 shows the the computed and analytic behaviour of $s$ as a
function of $t $ for the black hole arising in the classical
shockwave problem, demonstrating the stability and accuracy of the
algorithm.
As expected the behaviour is asymptotically linear.

When the same algorithm is applied to the solution of the semi
classical equations significantly different behaviour is encountered.
As can be seen in Figure~5, $s(t)$ is definitely non linear and
appears to be either bounded above, or at most logarithmic.
This behaviour is also observed for other values of the parameters. We
take this as an indication that the singularity does not remain in a
bounded region, like in a classical black hole, but spreads out to
infinity as a thunderbolt.

The decoupled ghost theory is also difficult to treat analytically
and so we resorted to numerical computation for this theory as well.
One of the arguments that were advanced for this theory was that
coupling the ghosts to a different metric would mean that a black hole
wouldn't radiate a negative energy flux if the number $N$ of scalars
was less than 24.  We therefore tried $N=12$, $\kappa =-1$.  However
the numerical results shown in figures~6 and 7 are radically
different.
We interpret them as appearing to indicate a
black hole that is growing in size with the apparent horizon moving
out.
Presumably this implies that the energy flux of the outgoing radiation
is negative.  We expect this to be true in any of the four theories if
$\kappa $ is negative.  We therefore calculated the more physically
reasonable case with $\kappa=+1$. This was similar to the original
CGHS theory.
Figure~8 shows $\arctan R$ as a function of $x^\pm$.
As in the earlier cases there is a singularity which is located
asymptotically at $x^- = const.$ as $x^+\rightarrow\infty$.
The apparent horizon lies to the past of the singularity and appears
tobe asymptotic to it.
The behaviour of $t(s)$ shown in figure~9 is also similar.
We had to integrate a lot further in this case but again it appeared
to be bounded indicating a thunderbolt singularity.

\beginsection 5. Conclusions

If an evaporating black hole does not settle down to a stable remnant,
any attempt to describe it by a Lorentz metric must have either a
naked singularity or a thunderbolt. We have studied four toy two
dimensional models of black hole formation and evaporation. In the two
theories whose Lagrangians had extra symmetry, the conformally
invariant and conserved current theories, it was possible to write the
general solution in terms of new variables, $\chi $ and $\Omega $. The
solution was non singular in terms of these variables, but it had a
naked singularity in terms of the physical variables, $\phi $ and
$\rho $. One might expect the semi classical approximation to break
down at this singularity and not to determine the fields beyond the
Cauchy horizon that starts at the point where the singularity first
becomes visible from infinity.

In the case of the two Lagrangians with extra symmetry, the Cauchy
horizon was regular when approached from below. However we suspect
that this will not be the case for more general Lagrangians: we expect
the singularity will be a spacelike or null thunderbolt that spreads
out to infinity and means that spacetime can be evolved only a finite
retarded time according to the semi classical equations. This
expectation is strengthened by the numerical calculations we have done
for two Lagrangians without special symmetries, the original model
proposed by CGHS, and the decoupled ghost modification proposed by
Strominger. The results point to thunderbolts in both cases if $
\kappa $ is positive, i.e.  if the number $N$ of minimal scalar fields
is greater than 24.

One can interpret these results as follows.  In two dimensions the
conservation and trace anomaly equations seem to imply that a solution
asymptotic to the linear dilaton will continue to radiate at a steady
rate.  Either the evolution of the solution will be cut off after a
finite time by a thunderbolt or the Bondi mass will become negative
eventually.  In the latter case, one might expect the singularity to
change from being spacelike to timelike and naked on about the outgoing
null line on which the Bondi mass becomes negative.  The four theories
we have considered illustrate the two possibilities:  the two with
additional symmetry give naked singularities with the Bondi mass
becoming arbitrarily negative while the two more general theories give
thunderbolts.

We would expect a semi classical treatment in four dimensions (if it
is possible to incorporate back reaction consistently) would be similar
to the more general two dimensional theories and would also lead to
thunderbolts in general. In our opinion, the prediction of a
thunderbolt
would indicate not that spacetime came to an end when a black hole
evaporated, but that the semi classical approximation broke down at
the end point. We would expect a full quantum treatment would soften
the thunderbolt singularity into a burst of high energy particles. It
would be tempting to try to connect such events with the gamma ray
bursts that have been observed, but there is a problem with the
energies involved. There is no reason to expect the semi classical
approximation to break down until the horizon size becomes of the
order the effective Planck length. In the case of black holes without
a conserved charge, this will not happen until the black hole gets
down to the Planck mass, so there's far too little energy left to
explain the observed gamma ray bursts, specially if they are at
cosmological distances, as the observations seem to indicate. Black
holes with a conserved charge but in theories without a dilaton field
approach a zero temperature extreme state, so the semi classical
approximation shouldn't break down and there's no reason to expect a
thunderbolt. In theory with a dilaton field with the coupling to
gauge fields suggested by string theory, the semi classical
approximation can break down while the black hole still has a
macroscopic mass. However, the mass difference between the black hole
at this point and the zero temperature extreme black hole, which is
presumably the ground state with the given charge, is much less than
the Planck mass. In fact it is even less than one quantum at the
temperature of the black hole. So any thunderbolt predicted by the
semi classical approximation would have to be extremely mild and could
not account for the observed gamma ray bursts. If the universe does
contain black holes that are reaching the end points of their
evaporation, it seems they will do it without much display.

\vfill
\eject

\beginsection Appendix.  Numerical Methods

This would seem to be the first paper in this area to utilize explicit
numerical solutions of the field equations.
This is somewhat surprising for in two dimensions reliable accurate
numerical solutions can be obtained readily using even modest computer
workstations.
The purpose of this section is to explain in some detail how our
numerical solutions were obtained, so that our methods become
accessible to others.

The fields are functions on the $x^\pm$ plane.
We replace the plane by a two dimensional lattice with equal spacing
$h$ in the $x^+$ and $x^-$ directions.
Figure 10 shows a typical lattice cell.
The four corners are denoted $n$, $e$, $s$ and $w$, while the centre
is denoted $o$.
Let $y(x^+, x^-)$ be a function taking values in $R^n$ and let $y_n$,
$y_e$, $y_s$, $y_w$ and $y_o$ be the values at the corresponding grid
points.
We assume that the function $y$ is sufficiently regular that it can be
represented within the cell by a Taylor series with remainder term
$O(h^4)$.
It is then a routine exercise to verify the following relations:
$$
\eqalignno {
y_o = & \half\(y_w + y_e\)  + O(h^2), &(A1a)\cr
y_o = & \quarter\(y_n+y_e+y_s+y_w\) + O(h^2), &(A1b)\cr
}
$$
$$
\eqalignno {
(\partial_+ y)_o = &\frac{y_e - y_s}{h} + O(h), &(A2a)\cr
(\partial_+ y)_o = &\frac{y_e - y_s + y_n - y_w}{2h} + O(h^2),
&(A2b)\cr
}
$$
together with the obvious analogues for $(\partial_- y)_o$, and
$$
\(\partial_+\partial_- y\)_o =
\frac{(y_n - y_e + y_s - y_w)}{h^2} + O(h^2). \eqno(A3)
$$

All of the theories treated here have field equations of the form
$$
\partial_+\partial_-y = F(y, \partial_\pm y) \eqno(A4)
$$
where $y = (\rho, \phi)^T$ and $F$ is smooth.
Further, initial data is given on the initial surfaces $x^\pm =
x^\pm_o$.
If we discretize (A4) and the initial data according to the above
prescription then the paradigm problem is the following: given $y_s$,
$y_e$ and $y_w$, determine $y_n$.
Evaluating (A4) at the point $o$ and using the relation (A3) we
obtain
$$
\eqalignno {
y_n = &y_w + y_e - y_s + h^2 \(F(y, \partial_\pm y)\)_o + O(h^4)\cr
 = &y_w + y_e - y_s + h^2 F\(y_o, (\partial_\pm y)_o\) + O(h^4),
&(A5) \cr
}
$$
where the last transformation is a tautology.
Nevertheless (A5) is the basis for our numerical algorithm.
We propose to evaluate it iteratively twice, leaving $y_w$, $y_e$ and
$y_s$ unaltered, but replacing the arguments of the function $F$ by
approximate values.

For our first evaluation we use approximations (A1a), (A2a) finding
$$
y_n := y_e + y_w - y_s + h^2F\(\half(y_w + y_e), h^{-1}(y_e - y_s)\) +
O(h^3) . \eqno(A6)
$$
Since all of the explicit terms on the right hand side of the equation
are known  we may evaluate a trial approximation to $y_n$, and
we have used here an atom of PASCAL formalism, ``$:=$'', whereby the
evaluated right hand side of the equation is then assigned to the
labelled quantity on the left hand side.

We can however do significantly better than this.
For our second evaluation we use approximations (A1b), (A2b) finding
$$
y_n := y_e + y_w - y_s + h^2F\(\quarter(y_s + y_e + y_s + y_w), \half
h^{-
1}(y_n - y_w + y_e - y_s)\) +
O(h^4) . \eqno(A7)
$$
In principle we could repeat this, regarding it as an iterative
process for solving the nonlinear equation (A7) for $y_n$.
However subsequent corrections to $y_n$ are smaller than the
truncation error $O(h^4)$ inherent in the equation.
Although the improvement in the error bound of (A7) over (A6) may
look small it is essential.
In order to integrate the equations out to large $x^+$ the algorithm
has to be applied $\sim h^{-3}$ times, and the errors committed at
each stage are cumulative.

\vfill
\eject

\centerline{\bf Figure Captions}

\beginsection Figure 1.

This figure shows some features of the Bilal and Callan exact solution
for $a=1.0$, $\kappa = 1.0$ and $T = 4.0$. The four curves show the
positions in the $x^\pm$ plane of the singularity, the apparent
horizon, what Bilal and Callan call a ``horizon'' and the Cauchy
horizon.

\beginsection Figure 2.

The surface drawn is $z = \arctan R(x^+, x^-)$ where $R$ is the Ricci
curvature for a classical black hole with $a=0.9$.
The solution is not defined to the future (right) of the singularity,
and so $R$ is assigned a token value of $\infty$.
Also shown is the apparent singularity, where $\partial_+\phi$ changes
sign.

\beginsection Figure 3.

The surface drawn is $z = \arctan R(x^+, x^-)$ where $R$ is the Ricci
curvature for  the original CGHS theory with $a=0.9$, $\kappa = 0.5$,
$N=30$.
The solution is not defined to the future (right) of the singularity,
and so $R$ is assigned a token value of $\infty$.
Also shown is the apparent singularity, where $\partial_+\phi$ changes
sign.

\beginsection Figure 4.

The affine parameter distance $s$ along an ingoing null geodesic from
the
initial surface $x^- = x^-_o$ to the apparent horizon is plotted
against $t$, the affine parameter distance along the initial surface.
Both curves refer to a classical black hole with $a=0.9$.
The solid line was obtained by numerical integration, while the dashed
line was computed analytically from equations (4.3).
As $t\rightarrow\infty$ the relation becomes linear.

\beginsection Figure 5.

The affine parameter distance $s$ along an ingoing null geodesic from
the
initial surface $x^- = x^-_o$ to the apparent horizon is plotted
against $t$, the affine parameter distance along the initial surface.
The solid line was obtained by numerical integration of the original
CGHS theory, while the dashed line was computed for a classical black
hole with the same initial data.
In both cases $a=0.9$, and for the CGHS theory $\kappa = 0.8$, $N=30$.
As $t\rightarrow\infty$ $s$ appears to be bounded above.

\beginsection Figure 6.

The surface drawn is $z = \arctan R(x^+, x^-)$ where $R$ is the Ricci
curvature for  the decoupled ghosts theory with $a=0.9$, $\kappa =
-1$, $N=12$.
The solution is not defined to the future (right) of the singularity,
and so $R$ is assigned a token value of $\infty$.
Also shown is the apparent singularity, where $\partial_+\phi$ changes
sign.

\beginsection Figure 7.

The affine parameter distance $s$ along an ingoing null geodesic from
the
initial surface $x^- = x^-_o$ to the apparent horizon is plotted
against $t$, the affine parameter distance along the initial surface.
The solid line was obtained by numerical integration of the
decoupled ghosts theory, while the dashed line was computed for a
classical black hole with the same initial data.
In both cases $a=0.9$, and for the semi-classical theory $\kappa =
-1$, $N=12$.
As $t\rightarrow\infty$ $s$ appears to be unbounded above.

\beginsection Figure 8.

The surface drawn is $z = \arctan R(x^+, x^-)$ where $R$ is the Ricci
curvature for  the decoupled ghosts theory with $a=0.9$, $\kappa =1$,
$N=36$ and $x^+_o = 4$.
The solution is not defined to the future (right) of the singularity,
and so $R$ is assigned a token value of $\infty$.
Also shown is the apparent singularity, where $\partial_+\phi$ changes
sign.

\beginsection Figure 9.

The affine parameter distance $s$ along an ingoing null geodesic from
the
initial surface $x^- = x^-_o$ to the apparent horizon is plotted
against $t$, the affine parameter distance along the initial surface.
The solid line was obtained by numerical integration of the
decoupled ghosts theory, while the dashed line was computed for a
classical black hole with the same initial data.
In both cases $a=0.9$, and for the semi-classical theory $\kappa = 1$,
$N=36$, $x^+_o = 4$.
As $t\rightarrow\infty$ $s$ appears to be bounded above.

\beginsection Figure 10.

The computational grid in the $x^\pm$-plane.
The plane is replaced by a lattice with spacing $h$.
Given data at points $s$, $w$ and $e$, the numerical algorithm
estimates $\partial_+\partial_-y$ at the fictitious point $o$ and
hence the dependent variable $y$ at the new lattice point $n$.

\vfill
\eject

\centerline{\bf References}
\frenchspacing

\item{[1]} C.G. Callan, S.B. Giddings, J.A. Harvey and A. Strominger,
{\it Evanescent black holes\/}, Phys. Rev. {\bf D45}, R1005--9, 1992.
\item{[2]} A. Strominger, {\it Fadeev-Popov ghosts and $1+1$
dimensional black hole evaporation\/}, Santa Barbara preprint
{UCSBTH-92-18}, hepth@xxx/9205028, May 1992.
\item{[3]} S.P. de Alwis, {\it Quantization of a theory of 2d dilaton
gravity\/}, University of Colorado preprint {COLO-HEP-280},
hepth@xxx/9205069, May 1992.
\item{[4]} A. Bilal and C.G. Callan, {\it Liouville models of black
hole evaporation\/}, Princeton University preprint {PUPT-1320},
hepth@xxx/9205089, May 1992.
\item{[5]} J.G. Russo, L. Susskind and L. Thorlacius, {\it The
endpoint of Hawking radiation\/}, Stanford University preprint
{SU-ITP-92-17}, June 1992.

\bye